\documentclass{nature}

\usepackage{graphicx}
\makeatletter
\let\saved@includegraphics\includegraphics
\AtBeginDocument{\let\includegraphics\saved@includegraphics}
\renewenvironment*{figure}{\@float{figure}}{\end@float}
\makeatother

\usepackage[english]{babel}
\usepackage{bbold}
\usepackage{bbm}
\usepackage{latexsym,amsmath,verbatim}
\usepackage{color}
\usepackage{rotating}
\usepackage{multirow}
\usepackage{color}
\usepackage{caption}

\usepackage{amssymb,amsfonts,amsmath}

\renewcommand{\(}{\left(}
\renewcommand{\)}{\right)}

\newcommand{\tr}[1]{\text{Tr}\(#1\)}
\renewcommand{\(}{\left(}
\renewcommand{\)}{\right)}

\usepackage{lineno}

\usepackage{xcolor}
\definecolor{RoyalBlue}{HTML}{4169e1}
\definecolor{ForestGreen}{HTML}{228b22}

\usepackage[colorlinks = true,
            linkcolor = blue,
            urlcolor  = blue,
            citecolor = blue,
            anchorcolor = blue]{hyperref}

\usepackage{comment}
\usepackage{subcaption}

\usepackage{todonotes}
\usepackage{lineno}

\definecolor{mypink2}{RGB}{219, 48, 122}

\title{Diversity of information pathways drives scaling and sparsity in real-world networks}

\author{Arsham Ghavasieh$^{1,2\ast}$, Manlio De Domenico$^{3,4,5\ast}$}

\begin{document}

\maketitle

\begin{affiliations}
\item Fondazione Bruno Kessler, Via Sommarive 18, 38123 Povo, Italy
\item Department of Physics, University of Trento, Via Sommarive 14, 38123 Povo, Italy
\item Department of Physics and Astronomy "Galileo Galilei", University of Padua, Via F. Marzolo 8, 315126 Padova, Italy
\item Padua Center for Network Medicine, University of Padua, Via F. Marzolo 8, 315126 Padova, Italy
\item Istituto Nazionale di Fisica Nucleare, Sez. Padova, Italy
\end{affiliations}

\vspace{-0.5truecm}
\noindent

$\ast$ Corresponding author: manlio.dedomenico@unipd.it,
aghavasieh@fbk.eu

\baselineskip24pt


\begin{abstract}
Empirical complex systems must differentially respond to external perturbations~\cite{eder2018biological,luppi2019consciousness} and, at the same time, internally distribute information to coordinate their components~\cite{kwapien2012physical,boguna2009navigability}. While networked backbones help with the latter~\cite{Watts1998}, they limit the components' individual degrees of freedom and reduce their collective dynamical range~\cite{ghavasieh2020statistical}. Here, we show that real-world networks are formed to optimize the gain in information flow and loss in response diversity.

Encoding network states as density matrices~\cite{de2016spectral,villegas2023laplacian}, we demonstrate that such a trade-off mathematically resembles the thermodynamic efficiency characterized by heat and work in physical systems. Our findings explain, analytically and numerically, the sparsity and the empirical scaling law observed in hundreds of real-world networks across multiple domains. We show, through numerical experiments in synthetic and biological networks, that ubiquitous topological features such as modularity and small-worldness emerge to optimize the above trade-off for middle- to large-scale information exchange between system's units. Our results highlight that the emergence of some of the most prevalent topological features of real-world networks have a thermodynamic origin.

\end{abstract}


Real-world networks are usually sparsely connected~\cite{busiello2017explorability} -- i.e., the number of existing links between units is much smaller than the one of potentially available links --  and exhibit peculiar topological properties like heterogeneous connectivity~\cite{barabasi1999emergence,broido2019scale,serafino2021true}, small-worldness~\cite{Watts1998}, modularity~\cite{newman2006modularity} and  hierarchy~\cite{ravasz2003hierarchical} as well as a balance between segregation and integration~\cite{tononi1994measure} or order and disorder~\cite{hidalgo2014information}. Several network growth mechanisms~\cite{albert2000topology,krapivsky2000connectivity,kumpula2007emergence,pacheco2006coevolution,garlaschelli2007self,dorogovtsev2002evolution,papadopoulos2012popularity,molkenthin2016scaling}, as well as methods not directly based on growth~\cite{holland1983stochastic,caldarelli2002scale,gallos2019propinquity}, have been proposed to replicate these features. However, a theoretical framework to explain why a certain transition from the disconnected state to a relatively stable wiring configuration is naturally favored is still lacking. This transition is observed in a variety of complex biological systems: communication among units -- that can be understood in terms of information exchange of chemical, electric or electrochemical signals, as well as binary packets or language  -- allow the system to start operating and functioning. For instance, human oral bacteria convey information within multispecies communities via signaling, such as adhesins and receptors, allowing for adherence and community development~\cite{Kolenbrander2002}. Once a fungal colony is established the cellular network use communication signals to regulate colony growth and development~\cite{leeder2011social,faust2012microbial}. Broadly speaking, information search and exchange plays an important role, still to be fully uncovered, in the formation, adaptation and evolution of living, synthetic or engineered complex systems~\cite{vergassola2007infotaxis}, which have to balance dynamic functions for a rapid response to internal and external perturbations with the energetic cost of the intervening actions~\cite{Nicholson2020}. 

In equilibrium and near-equilibrium statistical physics the above questions are naturally answered by fundamental principles, like the Gibbs entropy maximization or the free-energy minimization~\cite{jaynes1957information,jaynes1957informationII,lesar1989finite}. Conversely, for complex systems that are open and far from equilibrium, we lack an adequate theoretical framework to describe, explain and predict state transitions. Consequently, it is not clear how the aforementioned topological features observed in several real-world networks might emerge, and to which extent they are a causal byproduct of fundamental mechanisms related to how information between units is communicated and at which cost.

In the following, we will firstly introduce a theoretical framework for the analysis of network formation processes. Successively, we will show how this framework allows us to make predictions about specific topological features of complex networks, such as their connectedness and density, the coexistence of segregation and integration, as well as the coexistence of topological order and disorder responsible for the emergence of small-worldness. Finally, we compare our predictions for topological sparsity against hundreds of empirical networks, finding an excellent agreement between theoretical expectation and data.

\textbf{Network formation process.} Let $G$ be the structure of a complex system, modeled as a network of $N$ nodes and $|E|$ connections, respectively. The links between network units are often encoded into an adjacency matrix $\mathbf{A}$, where $A_{ij}=1$ if nodes $i$ and $j$ are connected, and $A_{ij}=0$ otherwise. System's units can exchange information in several ways, e.g. to create, destroy or efficiently use links in order to reach a stable functional regime, which is usually out of equilibrium because the system has to adapt to a changing environment. In the following we will generally refer to information exchange to characterize any type of signal that can be used for unit-unit communication from short- to long-range scales.

A variety of dynamical processes has been used to model the flow of information between the nodes~\cite{barzel2013universality,harush2017dynamic,hens2019spatiotemporal}. Yet, one of the simplest and most versatile dynamical process that can be used to model the flow of information through the networks is diffusion~\cite{Noh_RW_2004,barrat2008dynamical}, which is governed by the Laplacian matrix $\mathbf{L}=\mathbf{D}-\mathbf{A}$ where $\mathbf{D}$ is a diagonal matrix, with $D_{ii}=k_{i}$ --- i.e., degree--- being the connectivity of node $i$. The solution of the diffusion equation (See Methods) on top of networks can be understood in terms of a time evolution operator $e^{-\tau \mathbf{L}}$, where $\left(e^{-\tau \mathbf{L}}\right)_{ij}$ proxies the flow of information from node $j$ to $i$ and $\tau$ determines the propagation scale. For instance, when $\tau$ is small signals travel mostly between neighboring nodes, whereas large values of $\tau$ enable long-range communication.

The state of networks can be suitably encoded into density matrices, $\boldsymbol{\rho}_{\tau}=e^{-\tau \mathbf{L}}/Z$~\cite{de2016spectral,ghavasieh2020statistical}, with $Z=\tr{e^{-\tau \mathbf{L}}}$ being the partition function. Network density matrices have been used to tackle a broad range of problems, from classification of healthy and disease states~\cite{ghavasieh_SARSCOV2,ghavasieh_density_brain} and robustness analysis~\cite{ghavasieh_Structural_robustness,ghavasieh_functional_robustness} to dimensionality reduction of multilayer networks~\cite{de2015structural,ghavasieh2020enhancing} and renormalization group~\cite{villegas2023laplacian}. This relatively new framework is versatile because it is grounded in the physics of linear or non-linear response to the stochastic perturbations that occur at different locations and propagate throughout the links. Remarkably, the network Von Neumann entropy~\cite{de2016spectral} defined by $\mathcal{S}=-\tr{\boldsymbol{\rho}_{\tau}\log{\boldsymbol{\rho}_{\tau}}}$ measures how diverse the system responds to perturbations~\cite{ghavasieh2020statistical}, and the network free energy, $F = -\log{Z}/\tau $, is a measure of how fast the signal can transport between the nodes~\cite{ghavasieh2020enhancing}. This framework has been also used to characterize information cores in empirical networks~\cite{villegas2022laplacian} and multiscale brain functional connectivity~\cite{nicolini2020scale}.

Here, we characterize the network formation process as a transition between an initial state $G_{0}$ -- a collection of $N$ nodes with no links among them, $|E|=0$ -- and another network $G_{1}$, characterized by topological connectivity with certain properties. We show (See Methods) that this process can only lower the response diversity ($\delta \mathcal{S}\leq 0$), whereas it enhances the transport properties ($\delta F \geq 0$). To compare the two factors, we define a gain function $W = \delta F$ and a loss function $Q=\delta \mathcal{S}/\tau$, showing that $W \geq |Q|$ (See Methods), and their relative trade-off:

\begin{align}
    \eta = 1 - \frac{|Q|}{W}.
\end{align}

\begin{figure}[!h]
\centering
\includegraphics[width=.75\linewidth]{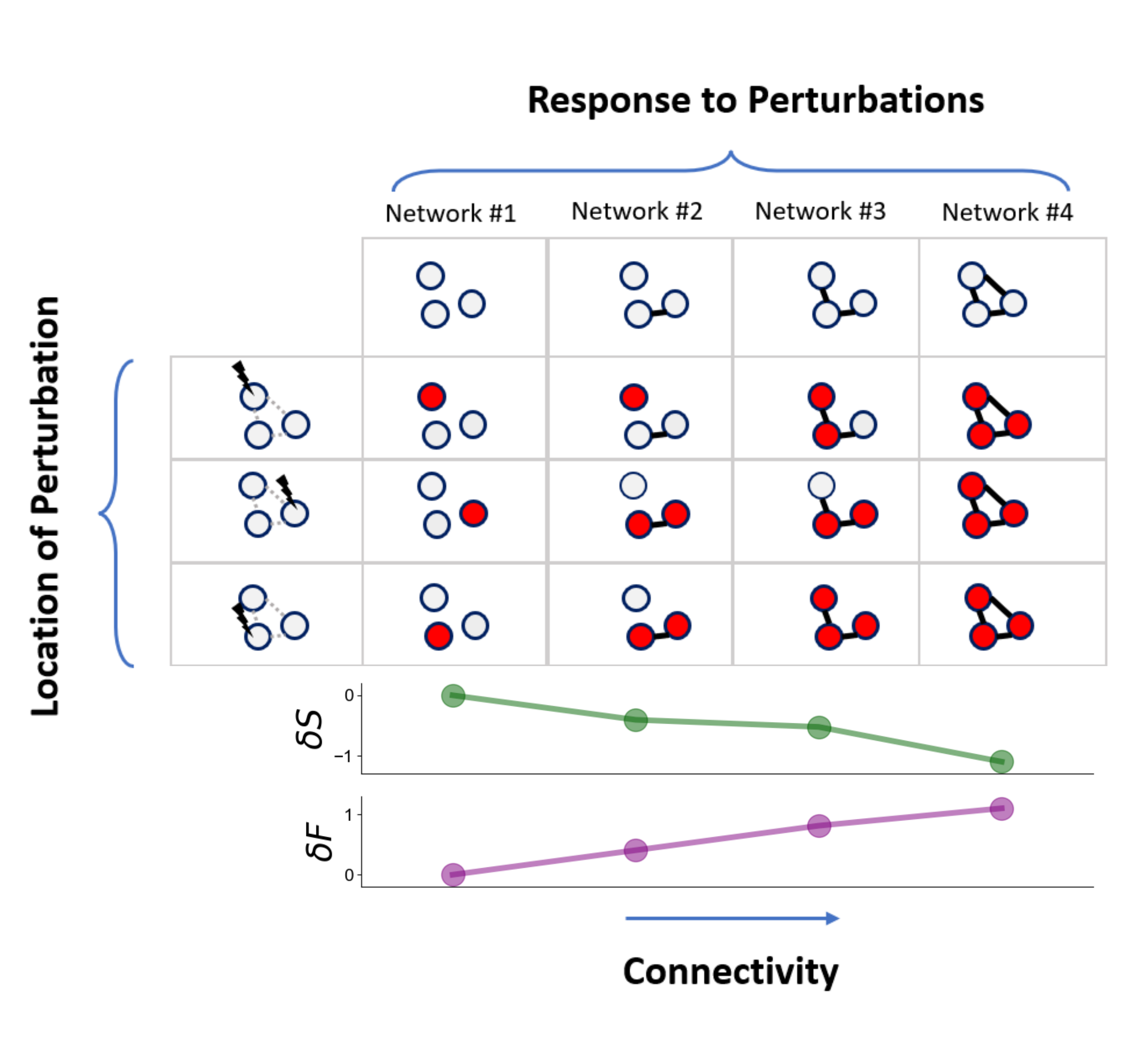}
\caption{\label{fig:schematic}\small{\textbf{Response diversity and information propagation.} $4$ simple graphs are considered, with different connectivities, and their response to different environmental perturbations is shown, with the perturbed nodes colored in red. For convenience, a simple dynamical rule is considered for the propagation of perturbations, where each perturbed node conveys the perturbation only to its neighbors. Trivially, connectivity enhances the flow of perturbations captured by the monotonic increase of $\delta F$ (See the text). For isolated nodes, the system's responses to different perturbations have no similarity (overlap). However, the overlap increases due to connectivity, reducing the diversity of responses. This is captured by the monotonically decreasing $\delta \mathcal{S}$ (See the text).}}
\end{figure}

Note that these quantities characterize the network formation process in terms of perturbation propagation and the diversity of systems' response. It is worth noticing that, from a mathematical perspective, they resemble the thermodynamic functions of heat $Q$ and work $W$, although such concepts cannot be directly extended to networks. In Fig.~\ref{fig:schematic}, an emblematic example is shown, where the response diversity and signal flow are compared among four graphs with different connectivities.

\textbf{$\eta$ and the signal propagation scale.} How deep the perturbations penetrate through networks depends on many factors, including the underlying topology, conductance, and the type of signals. Here, we use  $\tau$, the temporal parameter determining the scale of propagation, to characterize such a depth. For instance, at very small scales $(\tau \ll 1)$, the perturbation can not propagate through the system, whereas $\tau \gg 1$ let the perturbations flow between  topologically distant nodes.

Firstly, we Taylor expand the generalized heat and work functions while keeping only the leading terms. We show that at very small scales, where signals are contained locally $(\tau \approx 0)$, all network formation processes are indistinguishable, being characterized by $\eta=1$ (See Methods). Remarkably, the network topology becomes irrelevant to system's functionality during this regime. Secondly, we find that networks with fewer links, in the linear regime $\tau \ll 1$, exhibit higher values of $\eta$.

Finally, a second-order correction reveals the importance of connectivity: assuming that the number of links in a network scales with the number of nodes as $|E|=c N^{\gamma}$, where $c$ is a constant, we find that the optimal exponent which maximizes $\eta$ by $\partial_{\gamma}\eta = 0$ is obtained for $\gamma = 1$. Therefore, by requiring that $\eta$ is maximized by a network formation process, we naturally derive that the corresponding connectivity distribution must be sparse and scale with a specific exponent.

\begin{figure}[!h]
\centering
\includegraphics[width=\linewidth]{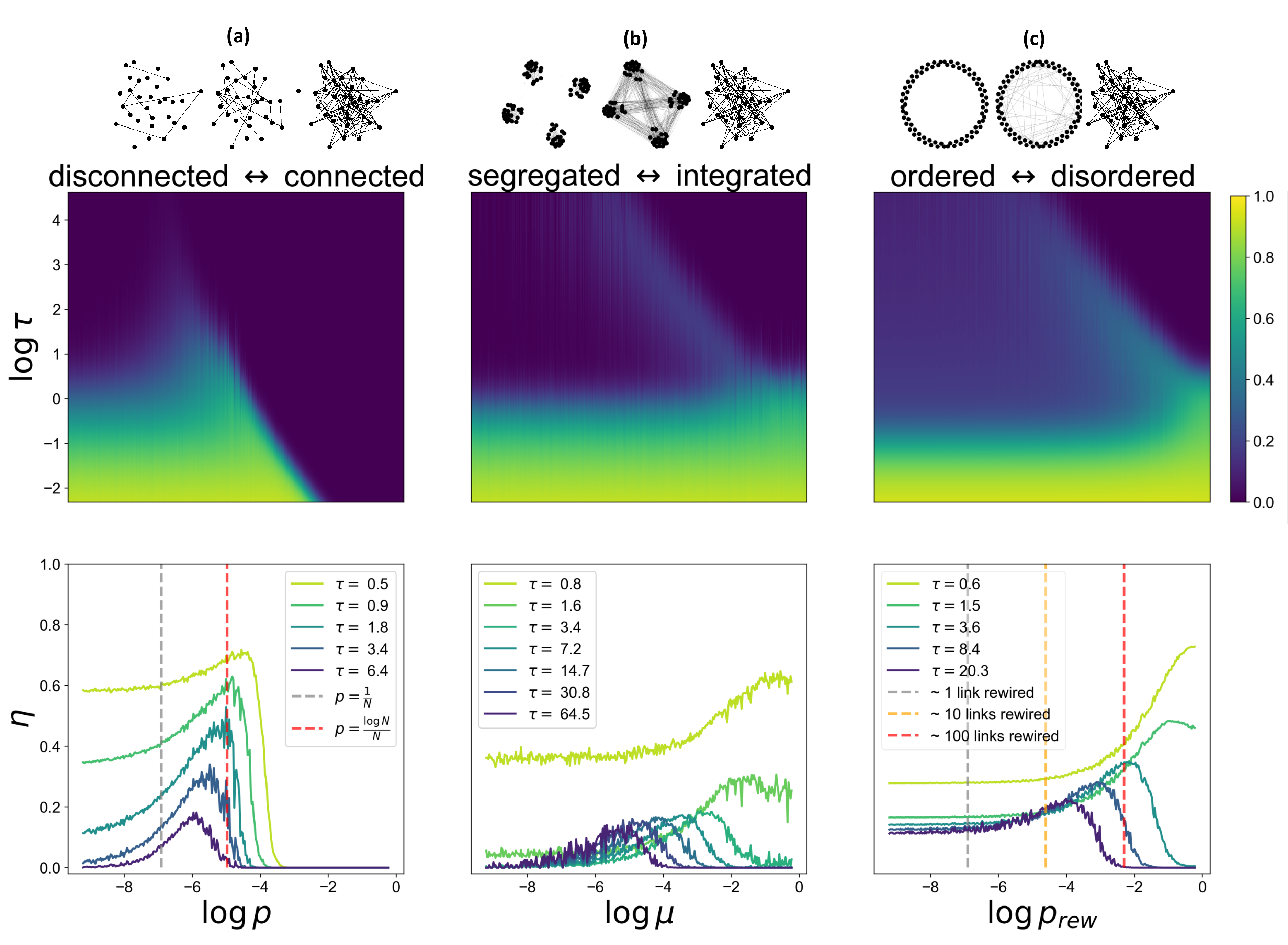}
\caption{\label{fig:synthetic}\small{\textbf{The effect of topological features on $\eta$.} (a) Connectivity versus disconnectivity, characterized by the emergence of a giant connected component, modeled by Erd\H{o}s-R\'enyi random networks for different wiring probability $p$. (b) Segregation versus integration, modeled by stochastic block networks for different mixing parameter $\mu$. (c) Order versus disorder, modeled by Watts-Strogatz small-world networks with degree $8$ and rewiring probability $p_{rew}$. In all three cases, different propagation scales $\tau$ are considered in the heatmap (top panels), while characteristic line curves have been separately shown for fixed values of $\tau$ (bottom panels) to better visualize transitions in $\eta$. In all cases, the size of synthetic networks is $N=10^3$.}}
\end{figure}

\begin{figure}[!h]
\centering
\includegraphics[width=.7 \linewidth]{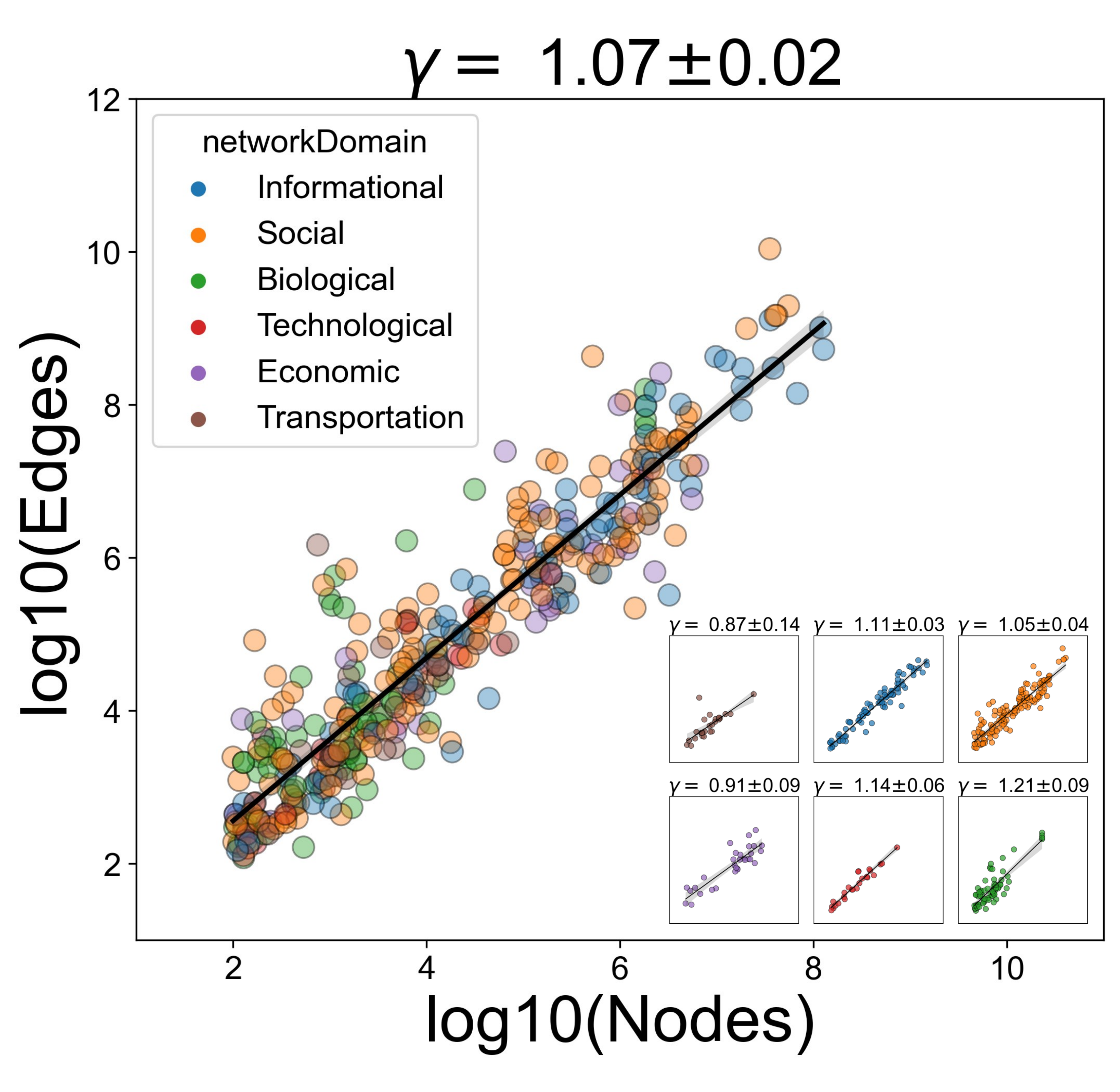}
\caption{\label{fig:scaling}\small{\textbf{Scaling in empirical networks.} The logarithm of the number of edges $|E|$ and nodes $N$ is represented for $543$ empirical networks, where the best fitting exponent is $\gamma = 1.07 \pm 0.02$. Similar analyses have been done for each network domain (informational, social, biological, technological, economic, and transportation.) separately and are represented in small blocks. The exponents are reported on top of each block.}}
\end{figure}

\textbf{The effect of topological features on $\eta$.} Here we study how some of the most prominent topological features of real-world networks affect $\eta$. More technically, these features include connectivity, integration versus segregation, and order versus disorder as captured by synthetic network (See Fig.~\ref{fig:synthetic}). 
First, we use Erd\H{o}s-R\'enyi (ER) networks~\cite{gilbert1959random,fienberg2012brief}, where the existence of links between every pair of nodes is independent and given by a connectivity probability $p$, to study the behavior of $\eta$ for varying $p$. Results indicate that being connected while sparse ensures high $\eta$: maximum $\eta$ is found between $p \propto N^{-1}$ and $p \propto N^{-1}\log{N}$ (See Fig.~\ref{fig:synthetic}). It is worth noting that in the thermodynamic limit $N \rightarrow \infty$, the difference between these two probabilities becomes negligible ($N^{-1} \approx N^{-1}\log{N}$), and the optimal probability can be approximated as $p\approx N^{-1}$. Therefore, given that in ER model the number of links is determined by the connectivity probability ($p$) as $|E|=p {N \choose 2}$, the numerical result validates our theoretical prediction of maximum $\eta$ at $|E|\propto N^{\gamma}$, with $\gamma \approx 1$.

Second, we consider a special type of stochastic block networks~\cite{peixoto2017nonparametric}--- i.e., models reproducing groups or communities of nodes that are randomly connected internally with probability $p_{in}$ and externally with probability $p_{out}$--- with $10$ communities and average degree $\Bar{k}=10$. This class of models is useful to investigate the mesoscale organization of complex networks. We vary the mixing parameter, defined as $\mu = \frac{k_{out}}{k_{in}+k_{out}}$, where $k_{in}$ is the average number of connections a node has with the nodes in its community, and $k_{out}$ is the average number of external connections. Values of $\mu \ll 1$ generate a highly segregated network with most connections existing only within a community, while $\mu \approx 1$ gives a network with no particular community structure, resembling ER models. In this case, at small scales ($\tau \ll 1$), the integrated communities exhibit the highest $\eta$, whereas at larger scales, the trade-off between integration and segregation -- with $\log{\mu}$ between $-6$ and $-2$ -- is favored. Therefore, our results confirm that in systems which are either too segregated or too integrated, the trade-off between gain in information flow and loss in response diversity is not optimal.

Third, we consider small-world networks--- i.e., models where nodes are initially connected in a lattice-like regular pattern, and then the links are shuffled with a given rewiring probability $p_{rew}$ to introduce topological shortcuts--- with average degree $\Bar{k}=8$. Results indicate that at small scales, $\eta$ is higher in disordered networks ($p_{rew}\approx 1$), while at larger scales, ordered structures ($p_{rew}\approx 0$) work better than disordered ones, with the optimal $\eta$ found in a middle regime between order and disorder exhibiting high small-worldness~\cite{Watts1998}. Once again, our framework provides an elegant explanation of why small-worldness is so ubiquitous in empirical systems.

\textbf{Scaling and sparsity of empirical networks.} We have analytically shown that if the number of links scales with the number of nodes as $|E|=c N^{\gamma}$, the optimal exponent which maximizes $\eta$ is given by $\gamma = 1$. This derivation implies that the real-world networks must be sparse in a specific way: their average degree must scale as $\Bar{k}=\frac{2|E|}{N}=\frac{2c N^{\gamma}}{N}= 2c$, and their connectivity must scale with the number of nodes. Here, we compare this theoretical expectation with empirical data.

We analyzed $543$ social, biological, informational, economic, transportation and technological networks from \href{https://icon.colorado.edu/#!/networks}{ICON}~\cite{index_of_complex_networks}, ranging in size from $N \approx 10^{2}$ to $N \approx 10^{8}$ nodes. Assuming a power-law ansatz between network size and the number of links, $|E|\propto N^{\gamma}$, we perform a linear regression in the log-log space to obtain the scaling exponent.
We find that the overall best fit is for $\gamma = 1.07 \pm 0.02$, while the proximity to $\gamma = 1$ result is relatively stable across domains (See Fig.~\ref{fig:scaling}), confirming our analytical and numerical prediction of $|E|\propto N^{1}$, with error bounds compatible with random topological fluctuations in network formation and noise in the data set.

Furthermore, we analyze hundreds of biological networks, from fungi to the human, C Elegans, and Ciona intestinalis connectomes (See Methods, Fig.~\ref{fig:fungal} and Fig.~\ref{fig:neural}). A direct comparison between these empirical networks and different ensembles of null models -- namely, Erd\H{o}s-R\'enyi (ER) and Configuration models (CM) -- (See Methods), shows that they are characterized by significantly higher $\eta$ at middle and large propagation scales, with respect to their randomized null models. This finding provides further evidence, supporting the numerical experiments of the previous section (See Fig.~\ref{fig:synthetic}), that complex systems must exhibit topological correlations to sustain their flow of information and, simultaneously, their pathway diversity at the macroscopic scales.

\textbf{Discussion.} While an interconnected structure is essential for information to flow between systems' units and for their coordination, it also limits the units' independence. This simple fact restricts their collective dynamical range, posing a significant challenge to complex systems that must diversely respond, and adapt, to unpredictable environments. Therefore, it is of paramount importance for real-world networks to exhibit topological features allowing them to ensure an adequate information flow among units, while simultaneously minimizing the cost of the diversity of available information pathways used for communication across multiple time scales.

Here, using network density matrices, we have characterized network formation as a physical process, quantifying the loss in response diversity, gain in information flow, and their trade-off. To this aim, we have shown that it is possible to introduce a new quantity ($\eta$) that, mathematically, resembles the thermodynamic efficiency related to heat and work. We analytically and numerically predict that networks must be sparse to optimize the trade-off, with their connectivity following a scaling law $|E|\propto N^{\gamma}$, with $\gamma \approx 1$. Our analysis of 543 empirical networks from biological, social, informational and transportation domains, reveals a scaling with exponent $\gamma \approx 1.07\pm 0.02$ that is strikingly compatible with our the theoretical expectation.

Moreover, our numerical experiments show that complex topological features like modularity and small-worldness can maximize $\eta$ when perturbations propagate beyond the first neighbors of nodes, enabling middle- to long-range communications between system's units. 
We validated these results with empirical data from hundreds of fungal and neural systems (See Methods, Fig.~\ref{fig:fungal} and Fig.~\ref{fig:neural}), while comparing against suitable null models that capture distinct topological features of the data. It demonstrates that topological correlations enable a complex network to maximize the trade-off and, therefore, play a crucial role in network formation and function.

Clearly, diffusion dynamics coupled with network topology is only a first step toward modeling the flow of information in empirical complex systems. This work opens the doors for future studies engaging nonlinear processes, especially reaction-diffusion, using suitable generalization of our framework. Furthermore, higher-order networks including multilayers~\cite{artime2022multilayer} and hypergraphs~\cite{battiston2021physics}, remain to be explored. Most importantly, relaxing some of our assumptions in the analytical calculation of the optimal exponent ($\gamma = 1$), like the mean-field approximation, can lead to even more precise predictions, spurring further interest for future developments of our approach.

Overall, our results are compatible with the compelling hypothesis that complex networks shape their topology to optimize the trade-off between maximally exchanging information among their units and guaranteeing an adequate response diversity for middle- to long-range signaling.

\textbf{Author Contributions} AG and MDD designed the study, performed the theoretical analysis and wrote the manuscript. AG performed the numerical experiments.

\textbf{Acknowledgements} MDD acknowledges financial support from the Human Frontier Science Program Organization (HFSP Ref. RGY0064/2022), from the University of Padua (PRD-BIRD 2022) and from the EU funding within the MUR PNRR “National Center for HPC, BIG DATA AND QUANTUM COMPUTING” (Project no. CN00000013 CN1).


\bibliographystyle{naturemag}

\begin{small}
\bibliography{biblio}
\end{small}

\textbf{Methods}

\textbf{Network density matrix.} Network density matrices have been recently generalized to non-linear dynamics~\cite{ghavasieh2022generalized}. However, here we focus on the diffusion dynamics governed by the combinatorial Laplacian, $\mathbf{L}=\mathbf{A}-\mathbf{D}$, where $\mathbf{A}$ and $\mathbf{D}$ are, respectively, the adjacency matrix ($A_{ij}=1$ indicates the existence of a link connecting node $i$ and $j$ and $A_{ij}=0$ indicates the absence of a link between the two) and degree diagonal matrix ($D_{ij}=\delta_{ij}k_{i}$ where $k_{i}$ is the degree of node $i$ and $\delta_{ij}=0$ if $i\neq j$ and $\delta_{ij}=1$ if $i=j$). The diffusion equation and its solution follows

\begin{align}\label{eq:continuous_dynamics}
    \partial_\tau|\psi_\tau \rangle = - \mathbf{L} |\psi_\tau \rangle, \qquad
    |\psi_{\tau}\rangle = e^{-\tau \mathbf{L}}|\psi_{0}\rangle.
\end{align}

From here, it is straightforward to obtain the density matrix $\boldsymbol{\rho}_{\tau}=e^{-\tau \mathbf{L}}/Z$~\cite{ghavasieh2020statistical}, with the partition function $Z=\tr{e^{-\tau \mathbf{L}}}$.

\textbf{Network counterpart of free energy.} In diffusion dynamics, the network partition function $Z$ measures the dynamical trapping~\cite{ghavasieh2020enhancing}--- i.e., the tendency of the network to block the flow of information locally. Let the eigenvalues of the Laplacian be $\lambda_{\ell}, \ell = 1, 2, ..., N$ for a network with $N$ nodes, the network partition function follows $Z = \sum\limits_{\ell=1}^{N} e^{-\tau \lambda_{\ell}}$. Since all the eigenvalues are non-negative, $\lambda_{\ell}\geq 0$, and at least one of the eigenvalues is $\lambda_{1}=0$, the partition function is bounded $1\leq Z\leq N$. The lower bound ($Z = 1$) is for a connected network at the very large propagation scales ($\tau \gg 1$). In contrast, the upper bound ($Z = N$) happens at the smallest scale, $\tau = 0$, or for any $\tau$ if the network is composed of isolated nodes--- because all eigenvalues of the Laplacian matrix are zero in that case.

The network free energy is inversely related to the partition function $F = - \log{Z_{\tau}}/\tau$. Accordingly, the network of isolated nodes has the lowest free energy $F_{min} = F_{iso}=-\log{N}/\tau$. Also, the minimum dynamical trapping ($Z_{\tau}=1$) gives the maximum free energy that is $F_{max} = 0$. Therefore, the network free energy is bounded as $0 \geq F \geq -\log{N}/\tau$. 

\textbf{Network counterpart of Von Neumann entropy.} The network Von Neumann entropy is given by $\mathcal{S}=-\tr{\boldsymbol{\rho_{\tau}}\log{\boldsymbol{\rho_{\tau}}}}$. Note that the eigenvalues of the density matrix are probabilities of a Boltzmann distribution ($e^{-\tau \lambda_{\ell}}/Z$) and the microcanonical ensemble gives the maximum entropy, where all eigenvalues are equal $\lambda_{\ell}=\lambda_{\ell^{\prime}}$. The only network satisfying this condition at $\tau > 0$ is the network of isolated nodes that has $\lambda_{\ell}=0, \ell=1,2,...,N$ and its Von Neumann entropy reads $S_{max}=S_{iso}=\log{N}$.

In contrast, in connected networks, only one eigenvalue of the Laplacian is $0$, and the rest are positive. Therefore, at large scales, all probabilities correspond to the non-zero eigenvalues and decay exponentially ($e^{-\tau \lambda_{\ell}}$), except for the first one that does not decay since $\lambda_{1} = 0$. This will lead to the minimum entropy $S_{min}=0$. 

Therefore, the network Von Neumann entropy is bounded as $0\leq \mathcal{S}\leq \log{N}$.

\textbf{Network counterpart of internal energy.} Because the network density matrix, in case of diffusion dynamics, resembles the Gibbs state, network internal energy can be obtained from the partition function as follows:

\begin{align}
    U= -\partial_{\tau}\log{Z}=\frac{1}{Z}\sum\limits_{\ell=1}^{N}\lambda_{\ell}e^{-\tau \lambda_{\ell}}.
\end{align}

Note that the summation of eigenvalues of the Laplacian matrix is twice the number of links $\tr{\mathbf{L}}=\sum\limits_{\ell=1}^{N}\lambda_{\ell}= 2|E|$, where $|E|$ is the number of links. Therefore, at $\tau = 0$, the network internal energy reads $U = \frac{1}{Z_{\tau}}\sum\limits_{\ell=1}^{N}\lambda_{\ell}=\frac{2|E|}{N}$. 

Also, it is worth mentioning that for a network of isolated nodes, $U_{iso} =0$, for all values of $\tau$, since all eigenvalues are $0$. 

Finally, the network's internal energy can be obtained from its free energy and entropy, as
\begin{align}
    U = F + \mathcal{S}/\tau
\end{align}

\textbf{Network formation as a physical process.} Network formation comes with an entropic cost since the entropy of isolated nodes is the maximum $\mathcal{S}_{iso}=\log{N}$. The cost is given by $\delta \mathcal{S}=\mathcal{S}-\mathcal{S}_{iso}$. Divided by the propagation scale $\tau$, it resembles heat dissipation in a thermodynamic process:
\begin{align}
    Q = \frac{\mathcal{S}-\log{N}}{\tau} \leq 0.
\end{align}

Furthermore, the gain in information flow can be obtained from the alteration of network free energy $\delta F = F - F_{iso}$, where $F_{iso}=-\log{N}/\tau$ is the minimum possible network free energy, at the fixed $\tau$, resembling the work in thermodynamics

\begin{align}
    W = F + \log{N}/\tau \geq 0.
\end{align}

Since $W+Q = F+\mathcal{S}/\tau=U$ and $U\geq 0$, the summation of gain and loss is always non-negative in the network formation process--- while $Q<0$ the gain is always larger or equal in magnitude $W\geq|Q|$.

Therefore, it is straightforward to show that $0< W+Q < W$ and $0<\frac{W+Q}{W}<1$, leading to the relative trade-off between the gain and loss in the network formation process, bounded between $0$ and $1$, that reads
\begin{align}\label{eq:U_and_eta}
    \eta = \frac{W + Q}{W}=1-\frac{|Q|}{W}=\frac{U}{W}.
\end{align}


\subsection{Approximation of $\eta$.} Let $\lambda_{\ell}$ be the $\ell$-th eigenvalue of the Laplacian matrix and $\Bar{\lambda^{n}}=\sum\limits_{\ell=1}^{N}\lambda^{n}_{\ell}$. It is straightforward to write the Taylor series for the partition function $Z = N \sum\limits_{\ell=1}^{N} \sum\limits_{n=0}^{\infty} \frac{(-\tau)^{n}}{n!}\frac{\lambda_{\ell}^{n}}{N}$, that we approximate up to the third-order $Z \approx N (1 - \tau \frac{\Bar{\lambda}}{N} +\frac{\tau^{2}}{2} \frac{\Bar{\lambda^{2}}}{N} - \frac{\tau^{3}}{6}\frac{\Bar{\lambda^{3}}}{N})$. From here, it is straightforward to approximate the logarithm of partition function, $-\log{\frac{Z}{N}} \approx  \tau \frac{\Bar{\lambda}}{N} -\frac{\tau^{2}}{2} \frac{\Bar{\lambda^{2}}}{N} + \frac{\tau^{3}}{6}\frac{\Bar{\lambda^{3}}}{N}$, that we use to write the second-order approximation of $U$ and $W$:

\begin{align}
    U &= -\partial_{\tau}\log{Z}= -\partial_{\tau}\log{\frac{Z}{N}} \approx  \frac{\Bar{\lambda}}{N} -\tau \frac{\Bar{\lambda^{2}}}{N} + \frac{\tau^{2}}{2}\frac{\Bar{\lambda^{3}}}{N} \\
    W &= -\frac{1}{\tau}\log{\frac{Z}{N}} \approx \frac{\Bar{\lambda}}{N} - \frac{\tau}{2} \frac{\Bar{\lambda^{2}}}{N} + \frac{\tau^{2}}{6}\frac{\Bar{\lambda^{3}}}{N}
\end{align}

For convenience, we assume that $\lambda_{\ell}^{2}=(\frac{\Bar{\lambda}}{N}-\delta\lambda_{\ell})^{2}$ and a apply a mean-field approximation $\Bar{\lambda^{2}} \approx \sum\limits_{\ell=1}^{N} (\frac{\Bar{\lambda}}{N})^{2}+\delta\lambda_{\ell}^{2}-2\delta\lambda_{\ell}\frac{\Bar{\lambda}}{N} \approx \frac{(\Bar{\lambda})^{2}}{N}$. Similarly, we assume that $\lambda_{\ell}^{3}=(\frac{\Bar{\lambda}}{N}-\delta\lambda_{\ell})^{3}$ and, therefore, a mean-field approximation leads to $\bar{\lambda^{3}}\approx \frac{(\Bar{\lambda})^{3}}{N^{2}}$. Finally, we assume that the number of links scales with the number of nodes as $\Bar{\lambda}= 2|E| = cN^{\delta + 1}$. Therefore, we can rewrite $U$ and $W$ as:

\begin{align}
    U &\approx  c N^{\delta}( 1 -\tau c N^{\delta} + \frac{\tau^{2}}{2}c^{2}N^{2\delta}) \\
    W &\approx cN^{\delta}(1-\frac{\tau}{2} cN^{\delta}+\frac{\tau^{2}}{6}c^{2}N^{2\delta}).
\end{align}

Finally, we expand the $W$ in the denominator of $\eta$ in Eq.~\ref{eq:U_and_eta} as:

\begin{align}
    \frac{1}{W} &= \frac{1}{cN^{\delta}}\frac{1}{1-\frac{\tau}{2} cN^{\delta}+\frac{\tau^{2}}{6}c^{2}N^{2\delta}} \nonumber \\
    &\approx \frac{1}{cN^{\delta}}(1 +\frac{\tau}{2} cN^{\delta}-\frac{\tau^{2}}{6}c^{2}N^{2\delta} + (\frac{\tau}{2} cN^{\delta}-\frac{\tau^{2}}{6}c^{2}N^{2\delta})^{2} )\nonumber \\
    &\approx \frac{1}{cN^{\delta}}(1 +\frac{\tau}{2} cN^{\delta}-\frac{\tau^{2}}{6}c^{2}N^{2\delta} + \frac{\tau^{2}}{4} c^{2}N^{2\delta}) \nonumber \\
    &= \frac{1}{cN^{\delta}}(1 +\frac{\tau}{2} cN^{\delta}+\frac{\tau^{2}}{12}c^{2}N^{2\delta}),
\end{align}

to reach the second-order approximation of the $\eta$:

\begin{align}
    \eta = \frac{U}{W}
    &\approx ( 1 -\tau c N^{\delta} + \frac{\tau^{2}}{2}c^{2}N^{2\delta})(1 +\frac{\tau}{2} cN^{\delta}+\frac{\tau^{2}}{12}c^{2}N^{2\delta}) \nonumber \\
    & \approx 1 - \frac{\tau}{2} c N^{\delta} +\frac{\tau^{2}}{12}c^{2}N^{2\delta} 
\end{align}

The first term $\eta \approx 1$ leads the behavior at extremely small $\tau\approx 0$. In that regime, the relative trade-off is equivalent for all network formation processes, indicating that all networks are indistinguishable and the topology is irrelevant to system function. 

The first order approximation reads $\eta \approx 1 - \frac{\tau c}{2}N^{\delta}$. Note that the second term is negative ($-\frac{\tau c}{2}N^{\delta}\leq 0$). Therefore, maximum $\eta$ corresponds to $\delta<0$, where $\displaystyle{\lim_{N \to \infty} N^{\delta}=0}$. In this regime, a disconnected network with as few as possible links $|E|$ is favored.

However, taking the derivative of the second-order approximation of $\eta$ with respect to the exponent $\delta$, reveals two optimal points.

\begin{align}\label{eq_derivative}
    \partial_{\delta} \eta &= -\frac{\tau c}{2}N^{\delta} [1 - \frac{\tau c}{3} N^{\delta}]\log{N} = 0.
\end{align}

One of the roots of Eq.~\ref{eq_derivative} requires $N^{\delta}=0$, favoring disconnected networks with $\delta < 0$, repeating the previously shown first-order approximation. However, the second root requires $N^{\delta}$ to remain a constant that does not scale up or down as $N \rightarrow \infty$, leading to the only possible answer $\delta = 0$. Note that in the text, the slope of Fig.~\ref{fig:scaling} is indicated as $\gamma\approx 1$ where $\gamma = \delta + 1$.

\textbf{Empirical $\eta$ profiles.} Here, we consider 270 biological systems representing networks of 6 species of fungi and slime molds: \emph{Physarum polycephalum} (Pp),  \emph{Phanerochaete velutina} (Pv), \emph{Resinicium bicolor} (Rb), \emph{Agrocybe gibberosa} (Ag), \emph{ Phallus impudicus} (Pi) and \emph{ Strophularia caerulea} Sc, introduced by Lee et al.~\cite{Lee2014}, where the weights of the links are proportional to the cord conductance for pairwise interactions. Furthermore, we consider the structural (DTI) and functional (fMRI) \emph{Rockland human connectome} data for 197 healthy subjects~\cite{Nooner2012_rockland}. We consider the structural (DTI) and functional (fMRI) \emph{Lausanne human connectome} data at four resolutions. We consider the structural \emph{Budapest human connectome} data, which is obtained from the consensus over  497 healthy subjects at three resolutions (20k, 200k, 1M), giving four networks, each representing one single type of connectivity between the brain areas: fiber count, fiber length, fractional anisotropy, and electrical connectivity~\cite{Budapest_connectome_v3}. We consider the connectome of a tadpole larva of \emph{Ciona intestinalis}~\cite{Ciona_connectome}. We consider the representations of \emph{Caenorhabditis elegans} (C. elegans) connectome, where links encode Send (S), Send-poly (Sp), Receive (R), Receive-poly (Rp) and Electric junction (Ej)~\cite{varshney2011structural,arnatkeviciute2018hub}.

We make an ER model and a configuration model (CM) for each empirical network as null statistical analysis models. Here, the ER null model with adjacency matrix $A^{(ER)}$ is a network of the same size (N) and the same number of links $|E|$ as the original network, where links encode a uniform connectivity probability (or, in case of weighted networks, density): $A_{ij}^{(ER)} = 2|E|/N(N-1)$. Instead, in the configuration model with adjacency matrix $A^{(CM)}$, the null model preserves the degree sequence, in addition to the number of links $|ER|$, with $A_{ij}^{(CM)}= k_{i}k_{j}/2|E|$ where $k_{i}=\sum\limits_{j=1}^{N}A_{ij}$ is the degree of node $i$.

See Fig.~\ref{fig:fungal} and Fig.~\ref{fig:neural} for a comparison of original networks with their ER and CM null models in fungal and neural systems. Consistently, randomness, represented by ER null models, provides the highest $\eta$ at small scales, $\tau\approx 0$--- with the only exception of fMRI correlation networks that show high similarity to the ER null model. At the same time, the degree distribution becomes essential for middle- to large-scale propagations. In some cases, the CM null model can explain the behavior of $\eta$, exhibiting no significant difference from the original network. However, topological features arising from degree-degree correlations lead to deviations from the CM null model in other networks. In conclusion, $\eta$ favors complex topological features at middle- to large-scale propagations, consistent with the synthetic network analysis in Fig.~\ref{fig:synthetic}.

\setcounter{figure}{0}
\makeatletter 
\renewcommand{\thefigure}{Extended Data~\@arabic\c@figure}
\makeatother

\begin{figure}[!h]
\centering
\includegraphics[width=.5\linewidth]{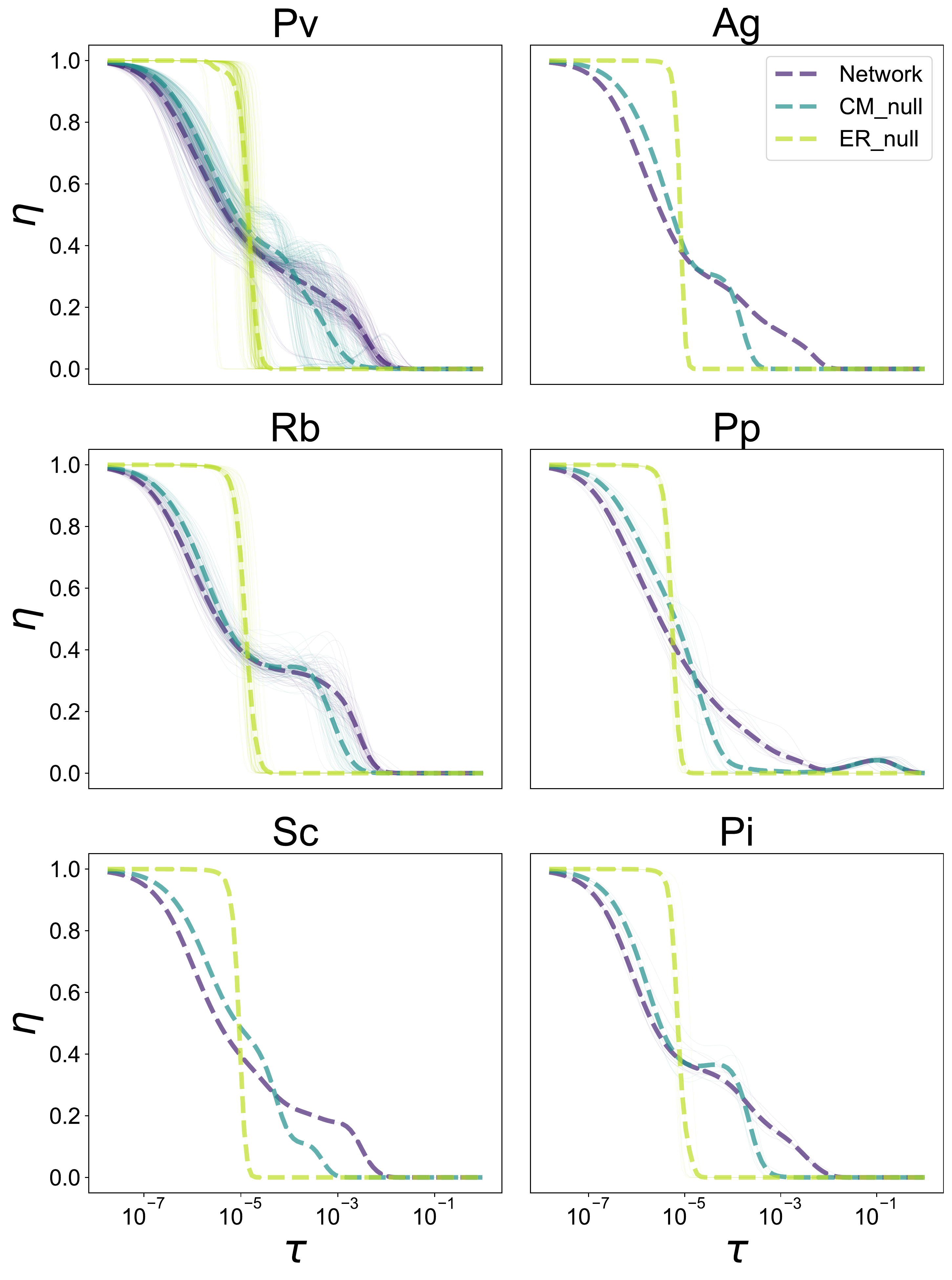}
\caption{\label{fig:fungal}\small{\textbf{Fungal networks compared with null models.} Six species of fungi are considered, including Pv, Ag, Rb, Pp, Sc, and Pi (see text for details). The dashed line shows the average $\eta$ for each species and its Erd\H{o}s-R\'enyi (ER) and Configuration Model (CM) surrogates.}}
\end{figure}

\begin{figure}[!h]
\centering
\includegraphics[width=\linewidth]{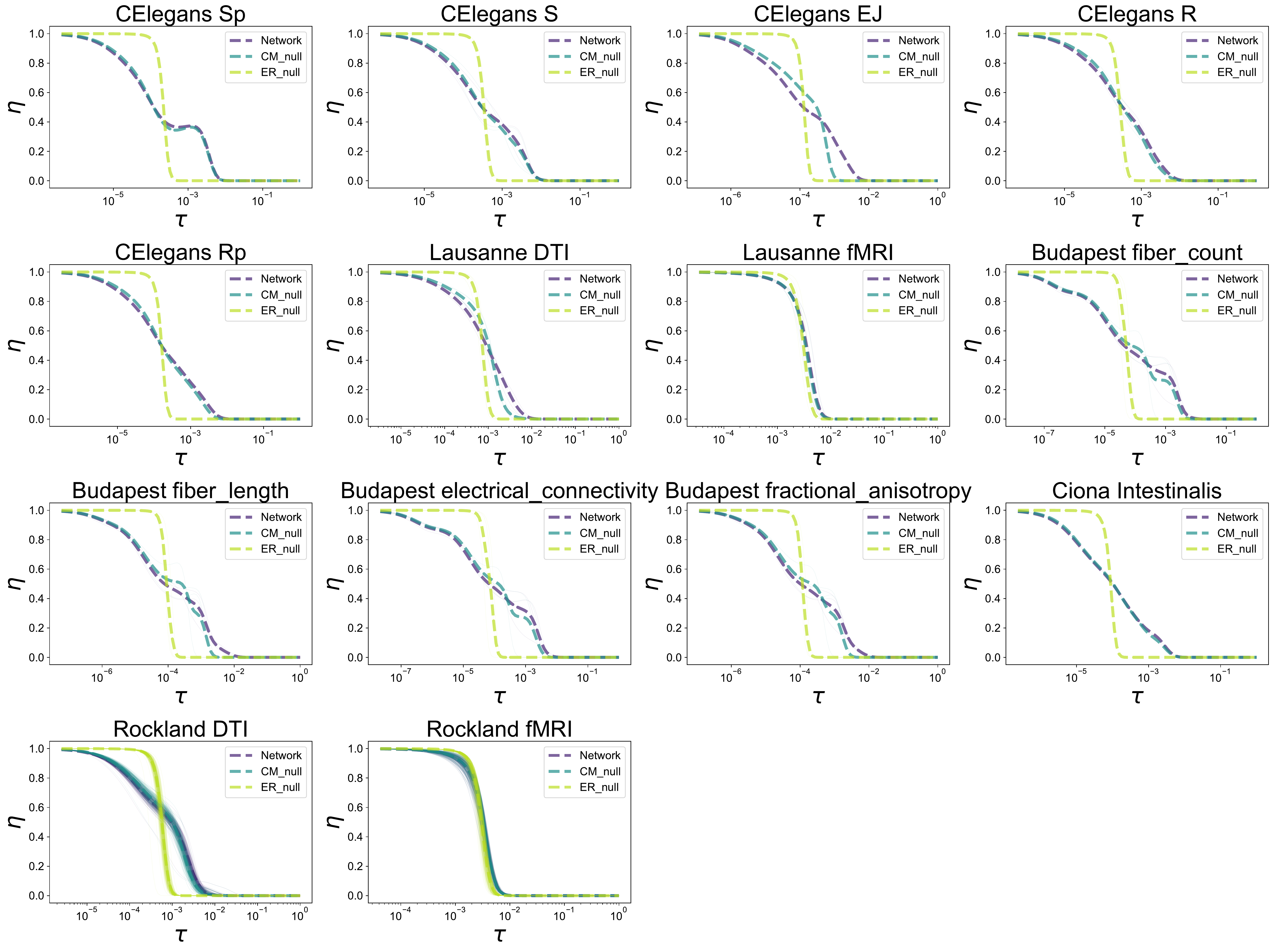}
\caption{\label{fig:neural}\small{\textbf{Neural networks compared with null models.} A wide range of connectomes are considered, including different types of connections in C. elegans, Ciona intestinalis, and structural and functional maps of the human connectome (See text for a description of each dataset). The dashed line shows the average $\eta$ for each species and its ER and CM null models, as in Fig.~\ref{fig:fungal}.}}
\end{figure}

\end{document}